\documentclass[12pt,a4paper,hypertex]{article}
\usepackage{jheppub_modified}

\allowdisplaybreaks[1]
\usepackage{amsmath,bm}
\usepackage{subfigure}

\title{Fast scrambling in holographic Einstein-Podolsky-Rosen pair}

\author{Keiju Murata}
\affiliation{Keio University, 4-1-1 Hiyoshi, Yokohama 223-8521, Japan}

\emailAdd{keiju@phys-h.keio.ac.jp}

\abstract{%
We demonstrate that a holographic model of the Einstein-Podolsky-Rosen pair exhibits fast scrambling. Strongly entangled quark and antiquark in $\mathcal{N}=4$ super Yang-Mills theory are considered. Their gravity dual is a fundamental string whose endpoints are uniformly accelerated in opposite direction. We slightly increase the acceleration of the endpoint and show that it quickly destroys the correlation between the quark and antiquark. The proper time scale of the destruction is $\tau_\ast\sim \beta \ln S$ where $\beta$ is the inverse Unruh temperature and $S$ is the entropy of the accelerating quark. We also evaluate the Lyapunov exponent from correlation function as $\lambda_L=2\pi/\beta$, which saturates the Lyapunov bound. Our results suggest that the fast scrambling or saturation of the Lyapunov bound do not directly imply the existence of an Einstein dual. When we slightly decrease the acceleration, the quark and antiquark are causally connected and an ``one-way traversable wormhole'' is created on the worldsheet. It causes the divergence of the correlation function between the quark and antiquark.
}

\begin{document}
\maketitle

\section{Introduction}

A local excitation of a quantum chaotic system spreads out over the entire system. 
This delocalization of the quantum information is called ``scrambling''.
It has been believed that the scrambling is related
to the quantum chaos or the butterfly effect:
initially similar states evolves completely different state at the late time.
The scrambling behaviour in strongly coupled systems
attracts much attention in the context of black hole physics or AdS/CFT correspondence~\cite{Maldacena:1997re,Gubser:1998bc,Witten:1998qj}.

It has been conjectured that
``Back holes are the fastest scramblers in nature''~\cite{Sekino:2008he}.
They demonstrated the delocalization
of local information on a black hole horizon
and estimated its time scale as $t_\ast\sim \beta \ln S$ where $\beta$ is the inverse Hawking temperature and $S$ is the Beckenstein-Hawking entropy.
This time scale is much quicker than that for usual quantum many body systems.
In Ref.~\cite{Shenker:2013pqa}, 
using the AdS/CFT correspondence, they developed a formulation to quantify the scrambling
in more concrete way. (See also Refs.~\cite{Shenker:2013yza,Roberts:2014isa,Leichenauer:2014nxa,Shenker:2014cwa,Jackson:2014nla,Polchinski:2015cea,Caputa:2015waa,Jahnke:2017iwi,Liu:2013iza})
They found that the scrambling behaviour can be read out from 
the mutual information or correlation function
between two boundaries of an eternal AdS black hole. 
Especially, the correlation function relates to 
the out-of-time-order correlator (OTOC).
In Ref.~\cite{Maldacena:2015waa}, it has been proposed that
the OTOC can be a measure of the quantum chaos and define
the ``quantum Lyapunov exponent'' from the OTOC.
It has also been conjectured that the Lyapunov exponent is bounded by the temperature:
$\lambda_L \leq 2\pi/\beta$.
Since the holographic calculation suggests that the ``Lyapunov exponent'' saturates the bound $\lambda_L=2\pi T$, the saturation of the bound is regarded as a sufficient condition that the field theory has its gravity dual.
The Sachdev-Ye-Kitaev model~\cite{Sachdev:1992fk,Kitaev-talk-KITP},
a quantum mechanics of Majorana fermions with all to all interactions,
is one of the examples which saturate the bound. 
This model is expected to describes a ``quantum'' black hole and, thus,
actively studied in recent years.

We propose one of the simplest models which exhibits the fast scrambling:
holographic Einstein-Podolsky-Rosen (EPR) pair~\cite{Xiao:2008nr,Jensen:2013ora}.
The conventional EPR pair is composed of two entangled electrons.
In the holographic EPR pair, on the other hand,
the quark and antiquark in $\mathcal{N}=4$ super Yang-Mills theory (SYM) are considered.
They are dual to a fundamental string hanging from the AdS boundary.
The string endpoints correspond to the quark and antiquark.
Endpoints are uniformly accelerated in opposite direction.
\begin{figure}
  \centering
  \subfigure[String profile]
 {\includegraphics[scale=0.45]{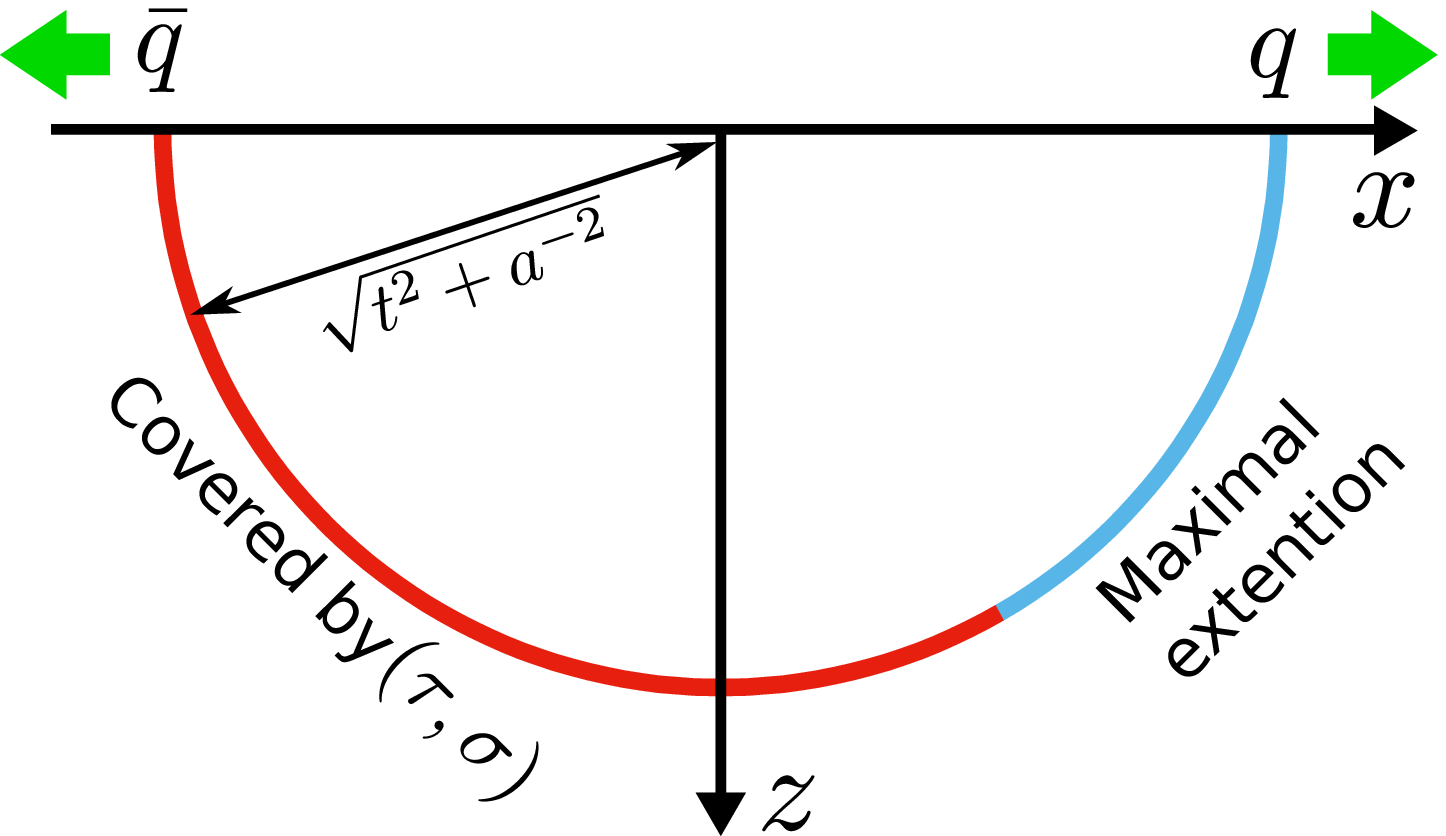}\label{semic}
  }
  \subfigure[Spacetime structure]
 {\includegraphics[scale=0.45]{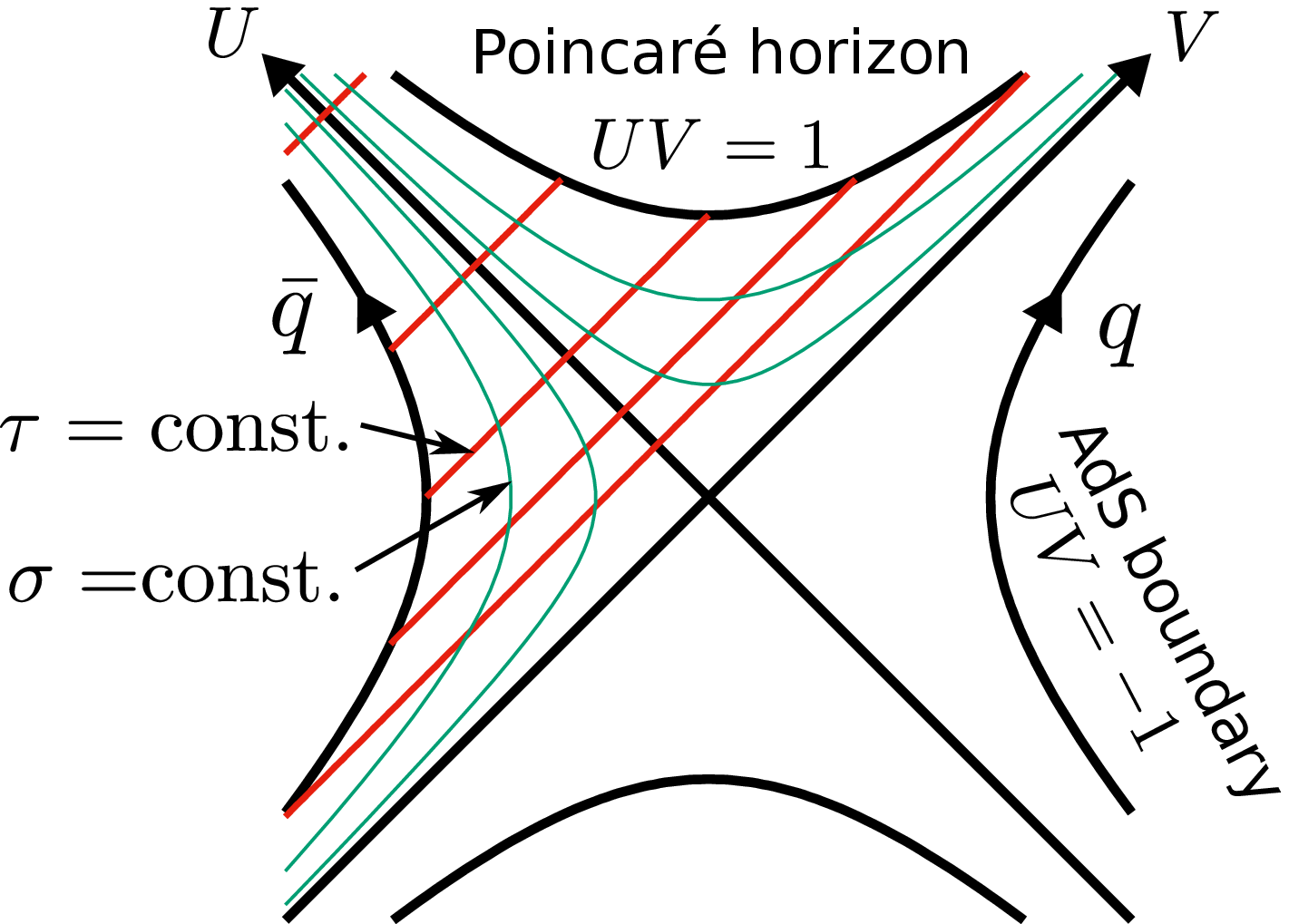}\label{AdS2BH}
  }
 \caption{
 (a) String profile of the holographic EPR pair for a fixed time slice.
 It is given by the semicircle whose radius is time-dependent.
 Endpoints of the string corresponds to the quark and antiquark.
 (b) Spacetime structure of the string worldsheet. Two time-like boundaries correspond to the quark and antiquark.
 They are causally disconnected but connected by the Einstein-Rosen bridge.
 }
\end{figure}
Fig.\ref{semic} shows the string profile of the holographic EPR pair.
The spacetime structure of the string worldsheet is shown in Fig.\ref{AdS2BH}.
From the figure, it is clear that 
the quark and antiquark are causally disconnected but connected
by a non-traversable Einstein-Rosen bridge.
Such a eternal black hole spacetime is
identified with strongly entangled state of two copy of CFTs~\cite{Maldacena:2001kr,Maldacena:2013xja}. 
Left and right quarks are causally disconnected but entangled. 
That is reason why
the accelerating quark-antiquark pair is called the holographic EPR pair.
In fact, Ref.\cite{Chen:2016xqz} showed the violation of the Bell inequality for the holographic EPR pair. 

We give a tiny perturbation to the antiquark (left CFT) following the idea of Ref.~\cite{Shenker:2013pqa}.
If the system exhibit chaos,
the quantum state of the antiquark is scrambled by the perturbation
and, as the result, the subtle relation 
between quark and antiquark is destroyed.
We measure the strength of the entanglement by the correlation function between quark and antiquark, which is equivalent to the OTOC.
We will see that the proper time scale of the decay of the correlation is given by
$\tau_\ast\sim \beta\ln S$ where $\beta$ is the inverse Unruh temperature and $S$ is the thermal entropy of the quark and gluons surrounding it.
The Lyapunov exponent read from the OTOC saturates the bound as $\lambda_L=2\pi/\beta$.
The dual picture of the EPR pair is the probe string in the fixed background.
In that sense, the dual theory is not Einstein gravity.
Nevertheless, we can see the fast scrambling.

The organization of this paper is as follows. 
In section.\ref{sec:EPR},
we introduce the holographic EPR pair as a solution in 
wide class of time dependant string solutions.
We also see that the spacetime structure of the string worldsheet
is same as an eternal AdS black hole.
In section.\ref{pEPR},
we explicitly construct the string solution with a perturbation.
Here, we consider the tiny change of the acceleration as the perturbation.
The geodesic distance 
between two endpoints of the perturbed string along the worldsheet is computed.
In section.\ref{sec:FS},
we compute the correlation function between the quark and antiquark.
It decays quickly and we find the fast scrambling result.
We also evaluate the Lyapunov exponent and find that it saturates the bound in Ref.\cite{Maldacena:2015waa}.
The effect of the decreased acceleration is also considered.
It changes the causal structure of the worldsheet drastically and creates the ``one-way traversable wormhole''.
We see the divergence of the correlation in this case.
In section.\ref{sec:two},
we consider the ``undo'' of the change of the acceleration:
First we change the acceleration as $a\to a'$ and, after a while, return it to a original value as $a'\to a$. We find that, even for this case, the EPR pair exhibits the fast scrambling. The final section is devoted to conclusion.

\section{Holographic EPR pair}
\label{sec:EPR}

\subsection{Exact solution for time dependant open strings}

We consider string dynamics in Poincare AdS$_5$ spacetime:
\begin{equation}
 ds^2=\frac{\ell^2}{z^2}(-dt^2+dz^2+dx^2+dy_1^2+dy_2^2)\ ,
 \label{AdS3}
\end{equation}
where $\ell$ is the AdS radius. We will take the unit of $\ell=1$ hereafter.
For simplicity, we will focus only on the string dynamics in the subspace spanned by $(t,z,x)$ 
although taking into account the dynamics in $(y_1,y_2)$ is straight forward.
The string dynamics is described by the Nambu-Goto action:
\begin{equation}
 S_\textrm{NG}=-\frac{1}{2\pi \alpha'}\int d^2\sigma \sqrt{-h}\ ,
\end{equation}
where $h$ is the determinant of the string induced metric.
In Ref.\cite{Mikhailov:2003er},
it has been shown that equations of motion for the string are solved by
\begin{equation}
 t=\frac{\dot{f}_t(\tau)}{\sigma}+f_t(\tau)\ ,\quad
 x=\frac{\dot{f}_x(\tau)}{\sigma}+f_x(\tau)\ ,\quad
 z=\frac{1}{\sigma}\ .
 \label{Miksol}
\end{equation}
We parametrized the string worldsheet by $(\tau,\sigma)$. Here, 
$f_t(\tau)$ and $f_x(\tau)$ are free functions satisfying
\begin{equation}
 -\dot{f}_t^2+\dot{f}_x^2=-1\ .
\label{fcond}
\end{equation}
The AdS boundary corresponds to $\sigma=\infty$.
The condition~(\ref{fcond}) implies that the worldsheet coordinate $\tau$
represents the proper time of the string endpoint.
The string induced metric for Eq.(\ref{Miksol}) is given by
\begin{equation}
 ds^2_h=-[\sigma^2-M(\tau)]d\tau^2+2d\tau d\sigma\ ,
  \label{induce}
\end{equation}
where we define 
\begin{equation}
 M(\tau)\equiv -\ddot{f}_t^2+\ddot{f}_x^2=\frac{\ddot{f}_x^2}{1+\dot{f}_x^2}\ .
\label{Mdef}
\end{equation}
This is the square of the acceleration of string endpoint.
The metric~(\ref{induce}) represents the two dimensional part of Vaidya-BTZ spacetime.

\subsection{Holographic EPR pair}

In Refs.~\cite{Hubeny:2014kma,Ghodrati:2015rta}, it has been realized that 
the holographic EPR pair solution is obtained as a special case of Eq.(\ref{Miksol}): 
\begin{equation}
 f_t(\tau)=\frac{1}{a}\sinh a \tau\ ,\quad
 f_x(\tau)=-\frac{1}{a}\cosh a \tau\ .
  \label{fsEPR}
\end{equation}
They satisfies Eq.(\ref{fcond}) and $M(\tau)=a^2$.
The string endpoint has a constant proper acceleration $a$.
For this solution, we can check the relation\footnote{
The AdS geometry~(\ref{AdS3}) has an isometry: 
$t'=2\ell a^{-1}t/\xi$, 
$x'=\ell (-t^2+x^2+z^2-a^{-2})/\xi$ 
and $z'=2\ell a^{-1}z/\xi$
where $\xi=-t^2+(x-a^{-1})^2+z^2$.
The EPR string solution is obtained by the coordinate transformation of a straight string sitting at $x'=0$.
}
\begin{equation}
 x^2+z^2=t^2+\frac{1}{a^2}\ .
  \label{xzt}
\end{equation}
For fixed $t$-slice, the string profile is semi-circle whose radius is given by $\sqrt{t^2+a^{-2}}$.
Fig.\ref{semic} shows string profile in $(x,z)$-plane for fixed $t$.
We will regard the endpoints at $x>0$ and $x<0$ as the quark and antiquark, respectively.
Our worldsheet coordinates $(\tau,\sigma)$ only cover the part of the semicircle shown by red curve.
We need an analytic continuation of the solution to cover the whole semicircle.
For the analytically continuation, we focus on the induced metric of the holographic EPR pair, which 
represents a static black hole spacetime in ingoing Eddington-Finkelstein coordinates:
\begin{equation}
 ds^2_h=-[\sigma^2-a^2]d\tau^2+2d\tau d\sigma\ .
  \label{induce2}
\end{equation}
For the maximal extension of the spacetime, we define null coordinates $(U,V)$ as
\begin{equation}
 U=e^{a\tau} \ , \quad V=-\frac{\sigma-a}{\sigma+a}e^{-a\tau} \quad \Longleftrightarrow\quad 
  \tau=\frac{1}{a}\ln U\ ,\quad \sigma=a\, \frac{1-UV}{1+UV}\ .
\label{UV_usigma}
\end{equation}
In terms of $(U,V)$-coordinates, the induced metric becomes
\begin{equation}
 ds^2_h=-\frac{4dUdV}{(1+UV)^2}\ .
  \label{induce3}
\end{equation}
This is nothing but AdS$_2$ spacetime.
The EPR string solution is written as
\begin{equation}
 t=\frac{1}{a}\frac{U+V}{1-UV}\ ,\quad x=-\frac{1}{a}\frac{U-V}{1-UV}\ , \quad z=\frac{1}{a}\frac{1+UV}{1-UV}\ .
  \label{txz_before}
\end{equation}
The $(U,V)$-coordinates cover the whole semicircle including light blue curve shown in Fig.\ref{semic}.

Fig.\ref{AdS2BH} shows the spacetime structure of the string worldsheet. The $(\tau,\sigma)$-coordinates cover only the top-left of the full spacetime.
While the induced metric is regular at $\sigma=0$ ($UV=1$), 
the string solution~(\ref{Miksol}) diverges at $\sigma=0$ in general.
The ``singularity'' $\sigma=0$ corresponds to the Poincar\'e horizon of the background AdS$_5$ spacetime.
Once we take the global coordinates for the target spacetime,
we can eliminate the coordinate singularity and extend the solution into $\sigma<0$.
In this paper, however, we focus only on the Poincar\'e patch of AdS$_5$ and regard $\sigma=a$
as the event horizon of the string worldsheet.\footnote{
This would be justified if we consider an AdS black holes with an infinitesimal mass as the target spacetime.
}

From Fig.\ref{semic}, it is clear that the left and right boundaries are connected by a non-traversable wormhole: There are two copy of  causally disconnected CFTs.
Such a spacetime is identified with the thermofield double state~\cite{Maldacena:2001kr,Maldacena:2013xja}:
\begin{equation}
 |\Psi\rangle = \frac{1}{\sqrt{Z}}\sum_n e^{-\beta E_n/2}|n\rangle_L |n\rangle_R \ .
\label{TFD}
\end{equation}
where $E_n$ is the energy eigenvalue and $|n\rangle_{L,R}$ is its eigenstate in left and right CFTs.
In terms of the thermofield double state,
the expectation value of an operator $\mathcal{O}_R$ in the right CFT is equivalent to the thermal expectation value: 
$\langle \Psi| \mathcal{O}_R |\Psi\rangle = Z^{-1}\textrm{tr}[e^{-\beta H_R}\mathcal{O}_R]$ where $H_R$ is the Hamiltonian of the right CFT. The thermal density matrix $\sim e^{-\beta H_R}$ originates from the entanglement of the left and right CFTs.
In the view of the thermofield double state,
the back hole entropy is obtained from the entanglement entropy.

\subsection{Thermodynamical variables}

It is known that
an accelerating point particle appears to be in a heat bath at the Unruh temperature~\cite{Unruh:1976db}:
\begin{equation}
 T=\beta^{-1}=\frac{a}{2\pi}\ .
  \label{Unruh}
\end{equation}
This coincides with the Hawking temperature obtained from the string induced metric~(\ref{induce2}).
It has also been confirmed that the quark and antiquark have the thermal entropy and energy:
\begin{equation}
 S=\frac{\sqrt{\lambda}}{3}\ ,\qquad E=\frac{4}{3}\sqrt{\lambda} T\ .
\label{ES}
\end{equation}
See the Supplemental Material of Ref.\cite{Jensen:2013ora} or Refs.\cite{Jensen:2013lxa,Casini:2011kv}
for the derivation of above expressions. To obtain above expressions, we 
change target space coordinates
in which the holographic EPR pair seems static.
From the onshell action of the string in the coordinate system,
entropy and energy can be computed.
The energy is not accompanied by the
translation of $t$ in Eq.(\ref{AdS3})
but by time translation in other coordinate system.

How do quarks gain non-zero thermal entropy? This is not a property of free SYM.
In the strongly coupled SYM plasma,
a quark attracts gluons and form a cloud of gluons around it~\cite{Jensen:2013ora}.
The cloud of gluons generates the non-zero entropy.
In that sense, quarks in SYM should be regarded as quasiparticles.

\section{Holographic EPR pair with a shock}
\label{pEPR}

\subsection{Perturbed holographic EPR string}
Since we know wide class of the exact solutions~(\ref{Miksol}),
we can give a perturbation by an analytic way as studied in Ref.\cite{Vegh:2015ska}.
As the perturbation,
we consider tiny change of the acceleration for the quark at $\tau=\tau_0$:
\begin{equation}
\begin{split}
&f_t(\tau)=
\begin{cases}
\frac{1}{a}\sinh a \tau & (\tau\leq \tau_0)\\
\frac{1}{a'}\sinh[a'\tau+c_1]+c_2\ & (\tau>\tau_0)
 \end{cases}\ ,\\
-&f_x(\tau)=
\begin{cases}
\frac{1}{a}\cosh a \tau & (\tau\leq \tau_0)\\
\frac{1}{a'}\cosh[a'\tau+c_1]+c_2'\ & (\tau>\tau_0)
 \end{cases}\ ,
\end{split}
\label{fspert}
\end{equation}
where $c_1$, $c_2$ and $c_2'$ are constants. From
continuity of above functions and their derivatives, these constants are chosen as
\begin{equation}
 c_1=-(a'-a)\tau_0\ ,\quad c_2=\left(\frac{1}{a}-\frac{1}{a'}\right)\sinh a\tau_0\ ,\quad
  c_2'=\left(\frac{1}{a}-\frac{1}{a'}\right)\cosh a\tau_0\ .
\label{cc}
\end{equation}
They are $C^1$-functions and their second derivatives are discontinuous.
One can also check that $-\dot{f}_t^2+\dot{f}_x^2=-1$ is satisfied.
Substituting above expressions into Eq.(\ref{Mdef}), we have
\begin{equation}
 M(\tau)=a^2+(a'{}^2-a^2)\theta(\tau-\tau_0)\ .
\end{equation}
Then, the induced metric
becomes 2d part of the Vaidya-BTZ spacetime with the ingoing shock.
We will refer the null surface $\tau=\tau_0$ as the shock surface.

Note that $\delta a= a'-a$ can be both positive and negative in our setup
because it is just a deviation of the acceleration caused by an external force.
On the other hand, in case of Vaidya spacetime,
we cannot reduce the black hole mass
unless we consider ``unphysical'' matter which violates the null energy condition.
For a while, we will focus on the case of $\delta a>0$. 
In section.\ref{deca}, we will consider the case of $\delta a<0$.

\begin{figure}
\begin{center}
\includegraphics[scale=0.6]{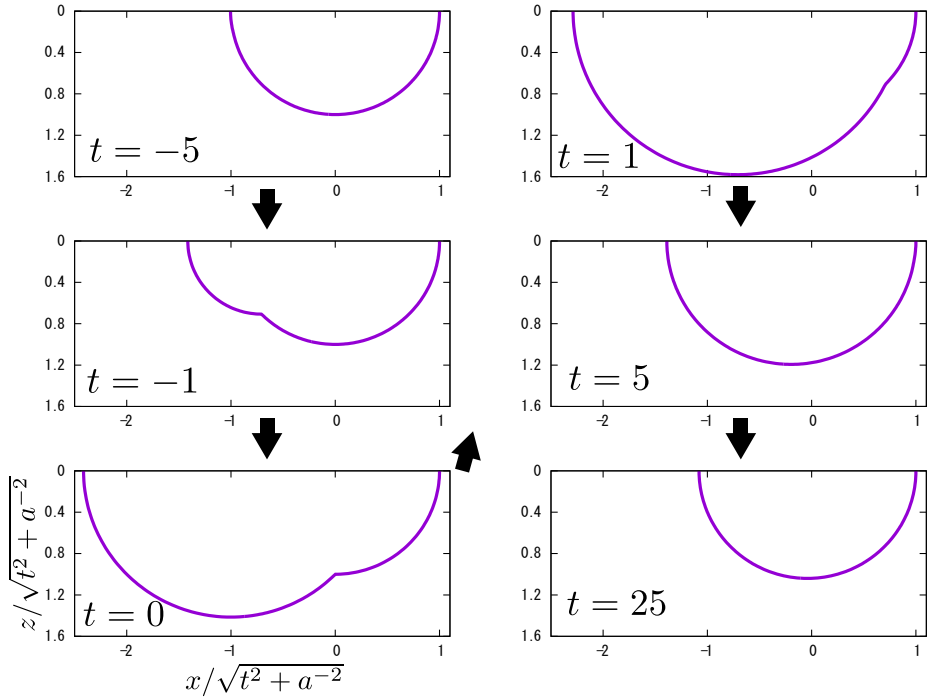}
\end{center}
 \caption{
 Time evolution of the string profile for several fixed time slices.
 Parameters are set as $\tau_0=-14.51$, $a=1$ and $a'=a+10^{-6}$. 
 Horizontal and vertical axes are proportional to $x$ and $z$ coordinates.
 They are normalized by $\sqrt{t^2+a^{-2}}$ to make the unperturbed string static.
}
\label{EPRev}
\end{figure}

Fig.\ref{EPRev} shows the time evolution of the perturbed string.
Since $\dot{f}_t$ and $\dot{f}_x$ have kinks at $\tau=\tau_0$,
the string profile specified by Eq.(\ref{Miksol}) also have
a kink.
As the parameter for this figure, we took $\tau_0=-14.51$, $a=1$ and $a'=a+10^{-6}$.
Especially, the change of the acceleration is really tiny: $\delta a =10^{-6}$.
Nevertheless, it causes a significant change in the profile of the string around at $t=0$.
This is a string realization of the phenomena found in BTZ black hole~\cite{Shenker:2013pqa}:
The early infalling quanta can create a strong shock wave by the effect of the blue shift at the white hole horizon.

Before the shock $\tau\leq \tau_0$, we use $(U,V)$-coordinates defined in Eq.(\ref{UV_usigma}).
After the shock, we have $M(\tau)=a'{}^2$.
We introduce the other double null coordinates $(U',V')$ as
\begin{equation}
 U'=e^{a'\tau} \ , \quad V'=-\frac{\sigma-a'}{\sigma+a'}e^{-a'\tau} \quad \Longleftrightarrow\quad 
  \tau=\frac{1}{a'}\ln U'\ ,\quad \sigma=a'\, \frac{1-U'V'}{1+U'V'}\ .
\label{UpVp_usigma}
\end{equation}
In terms of $(U',V')$-coordinates, the string solution is written as
\begin{equation}
\begin{split}
 &t=\frac{\delta a}{aa'} \sinh a \tau_0 +\frac{1}{a'}\frac{U'e^{-\delta a \tau_0}+V'e^{\delta a \tau_0}}{1-U'V'}\ ,\\
 &x=-\frac{\delta a}{aa'} \cosh a \tau_0 -\frac{1}{a'}\frac{U'e^{- \delta a \tau_0}-V'e^{\delta a \tau_0}}{1-U'V'}\ ,\quad
 z=\frac{1}{a'}\frac{1+U'V'}{1-U'V'}\ ,
\end{split}
\end{equation}
where $\delta a =a'-a$.

\begin{figure}
\begin{center}
\includegraphics[scale=0.7]{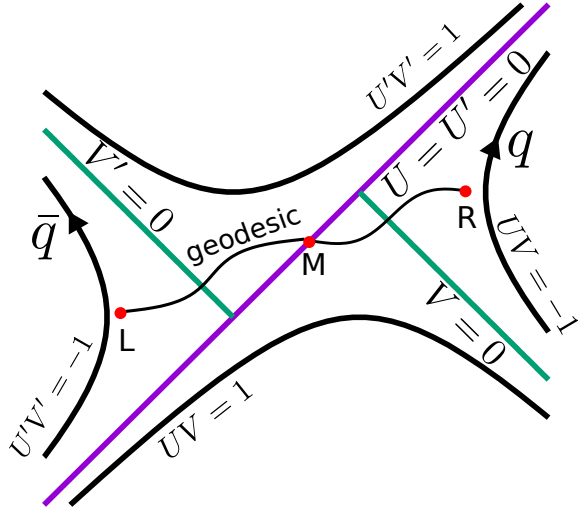}
\end{center}
\caption{
 Spacetime structure of the perturbed string worldsheet.
 There is the shift $\gamma$ in origins of $(U,V)$- and $(U',V')$-coordinates. 
}
\label{shock}
\end{figure}

Since coordinate $(U,V)$ and $(U',V')$ are defined individually in $\tau\leq \tau_0$ and $\tau>\tau_0$, 
we need to determine the matching condition for $(U,V)$ and $(U',V')$ at the shock surface.
In $(U,V)$- and $(U',V')$-coordinates, the shock surface is written as
$U=e^{a\tau_0}$ and $U'=e^{a'\tau_0}$ respectively.
Since the coordinates $(\tau,\sigma)$ are common in in Eqs.(\ref{UpVp_usigma}) and (\ref{UV_usigma}),
we can obtain the matching condition for $V$ and $V'$ as
\begin{equation}
 a\, \frac{1-UV}{1+UV}\bigg|_{\tau=\tau_0} = a'\, \frac{1-U'V'}{1+U'V'}\bigg|_{\tau=\tau_0}\ .
\end{equation}
Solving the equation with respect to $V'$, we have 
\begin{equation}
 V'=\frac{\delta a + (a+a')\,e^{a\tau_0} \, V}{\delta a \,e^{a\tau_0} \, V + (a+a')}e^{-a' \tau_0}\ .
\end{equation}
As mentioned in Ref.\cite{Shenker:2013pqa},
the matching condition becomes simple in the double scaling limit:
\begin{equation}
 \delta a \to 0\ ,\qquad \tau_0\to -\infty\ ,\qquad \gamma\equiv \frac{\delta a}{2a}e^{-a \tau_0}\textrm{ : fixed}\ .
\label{SSlimit}
\end{equation}
In this limit, the matching condition is written as
\begin{equation}
 V'=V+\gamma\ ,
 \label{Vshift}
\end{equation}
and the shock surface is given by $U=U'=0$.
Fig.\ref{shock} shows the causal structure of
the induced metric of the perturbed EPR string in the limit of Eq.(\ref{SSlimit}).

\subsection{Geodesic distance}

\begin{figure}
  \centering
 \subfigure[$\tau_0=-14.51$ ($\gamma=1$)]
 {\includegraphics[scale=0.25]{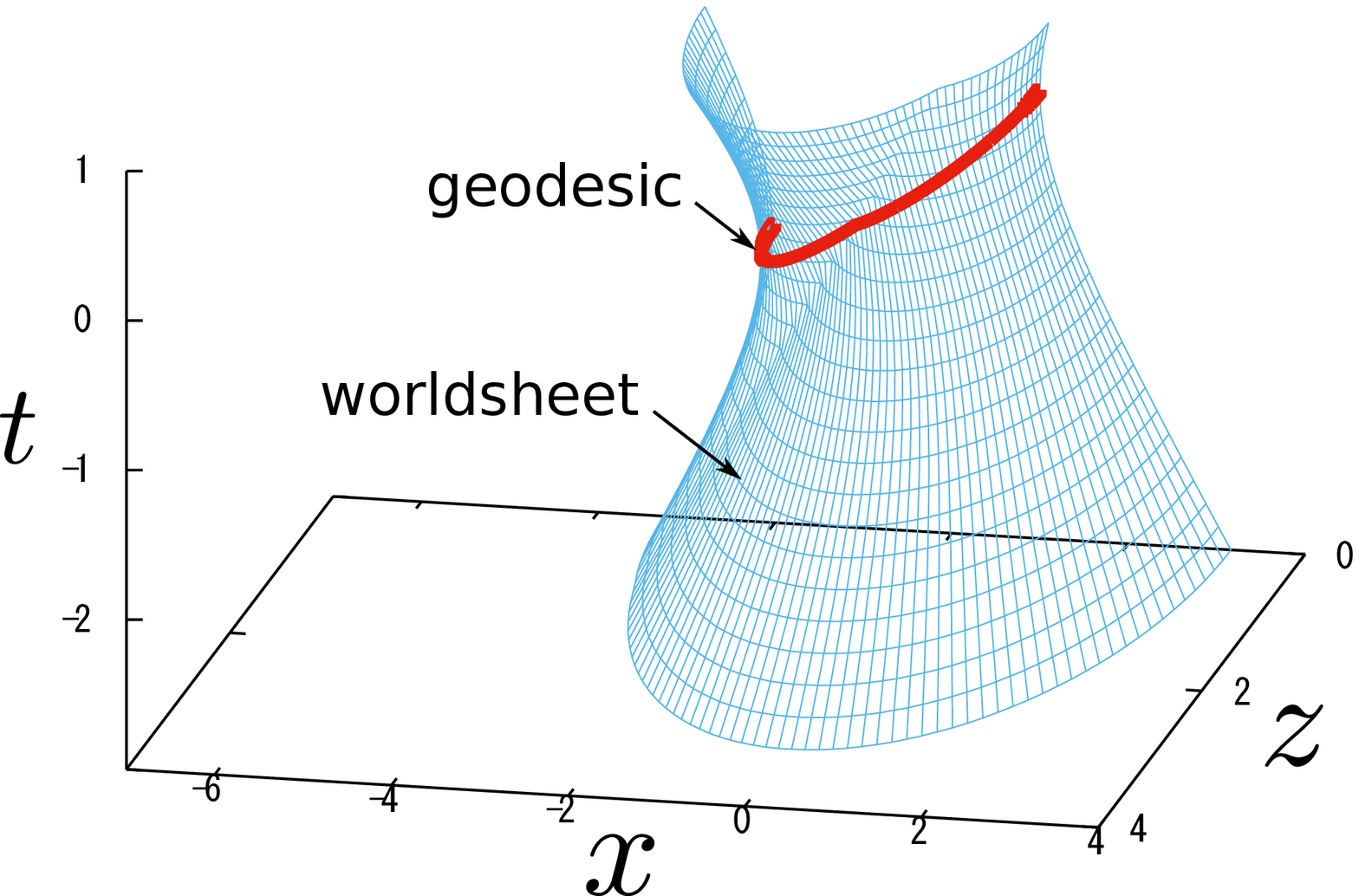}\label{g1}
  }
  \subfigure[$\tau_0=-15.20$ ($\gamma=2$)]
 {\includegraphics[scale=0.25]{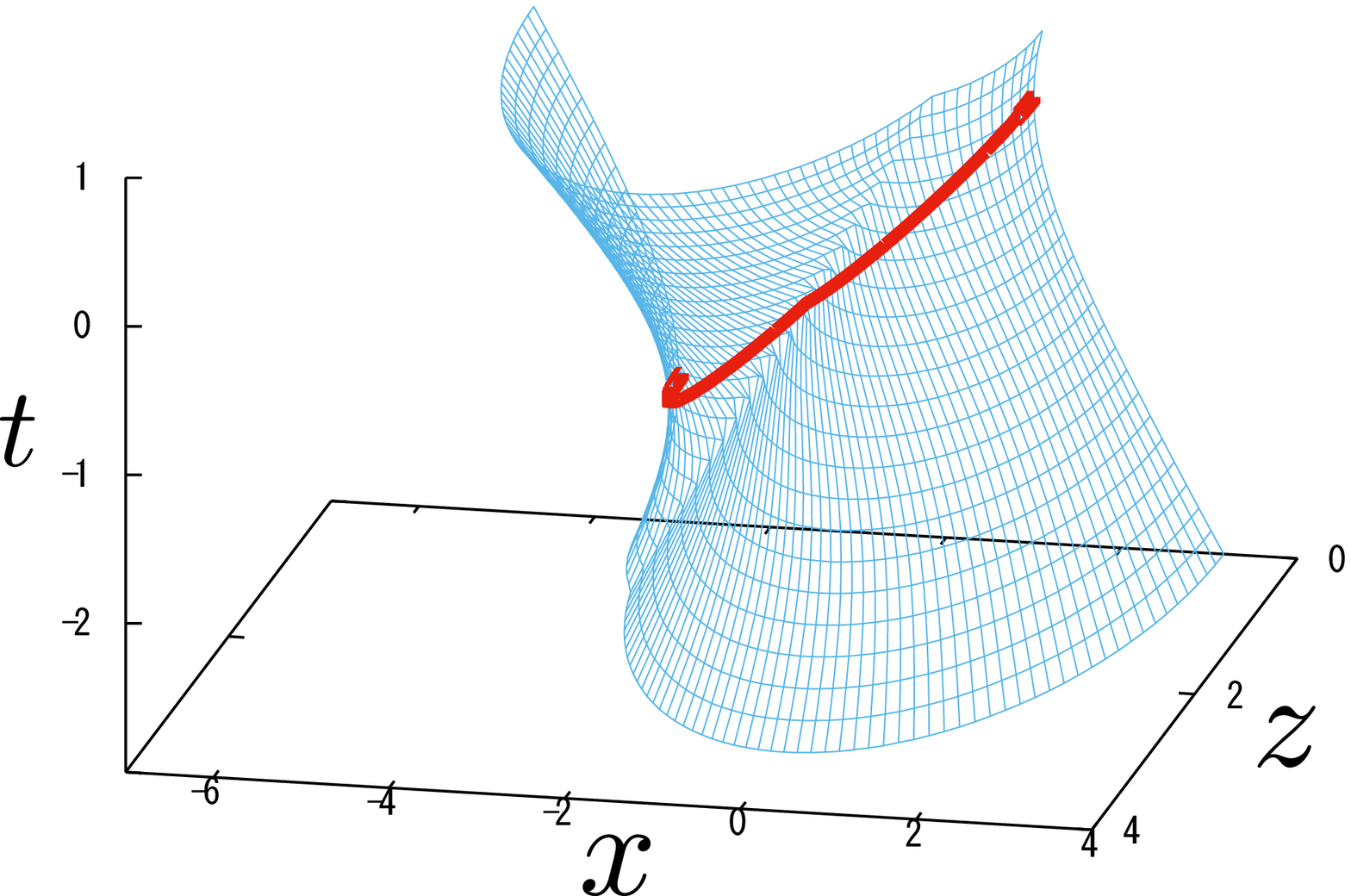}\label{g2}
 }
 \subfigure[$\tau_0=-15.61$ ($\gamma=3$)]
 {\includegraphics[scale=0.25]{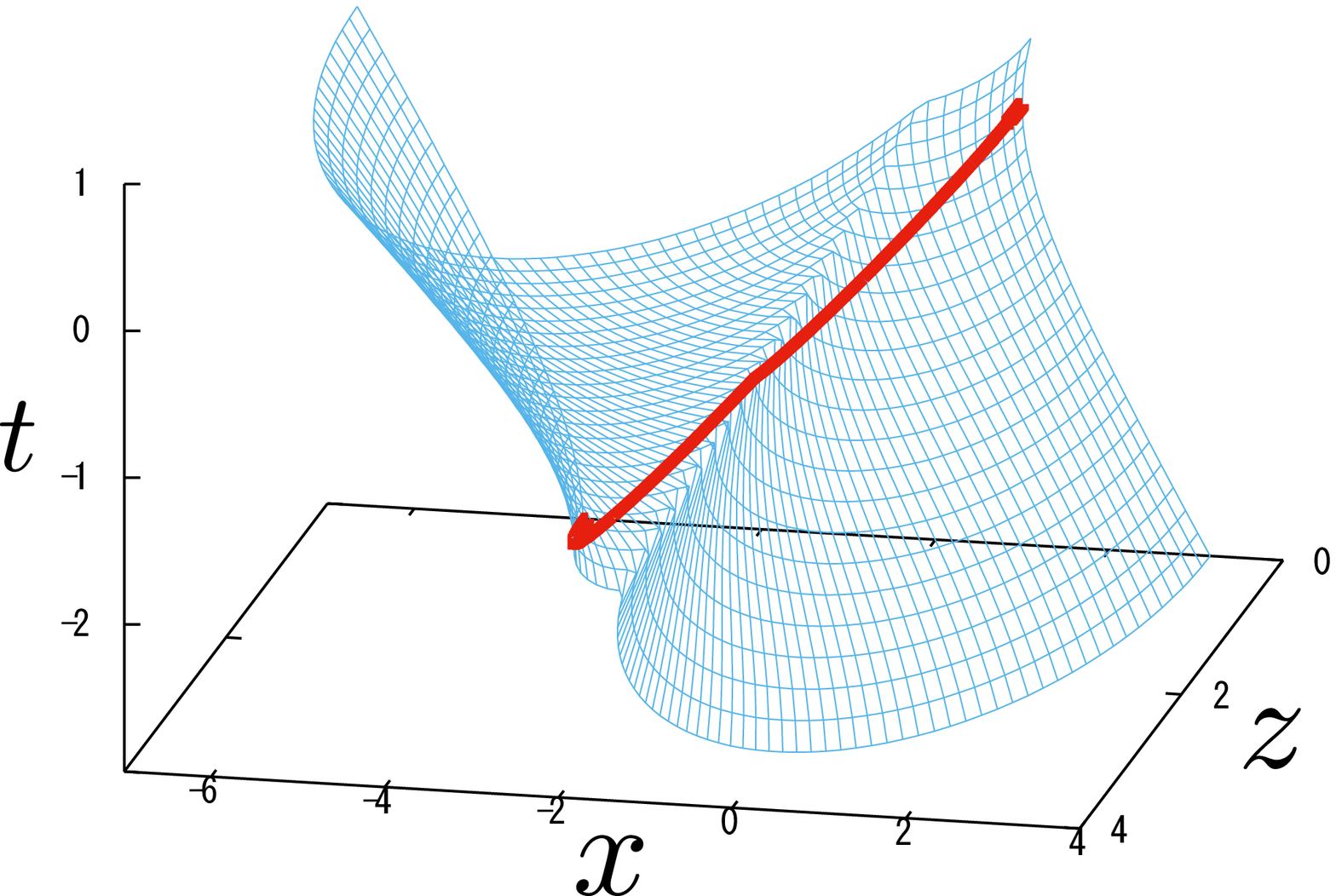}\label{g3}
  }
 \caption{
 String worldsheets and geodesics between boundaries at $\tau_R=0$ and $\tau_L=0$ 
 for $a=1$ and $a'=a+10^{-6}$.
 The proper times for the shock surface $\tau_0$
 is chosen so that $\gamma=1, 2$ and $3$ in Figs.(a), (b) and (c), respectively.  
 }
 \label{geodesic_worldsheet}
\end{figure}

For the purpose of estimating the correlator between quark and antiquark in next section, 
we compute the geodesic distance
between points in left and right boundaries of the  worldsheet.
We take the double scaling limit~(\ref{SSlimit}) to simplify following expressions.
In Fig.\ref{shock}, we show the schematic picture of the geodesic between points
L: $(U'_L,V'_L)$ and R: $(U_R,V_R)$.
The shock surface is located at $U=U'=0$.
Geodesics in left and right region connect at the point M: $(0,V_M)$ which is on the shock surface.
To regularize the distance, we put A and B slightly inside of AdS boundaries.
The induced metric of the perturbed EPR string is locally AdS$_2$ for $U\neq 0$.
Thus, we can embed the geometry of the string worldsheet
in $\bm{R}^{2,1}$ for $U<0$ and $U>0$ respectively.
For $U<0$, we define the coordinates of $\bm{R}^{2,1}$ as 
\begin{equation}
T_1=\frac{V+U}{1+UV}\ ,\quad
T_2=\frac{1-UV}{1+UV}\ ,\quad
X_1=\frac{V-U}{1+UV}\ ,
\label{R21emb}
\end{equation}
One can check $-T_1^2-T_2^2+X_1^2=-1$ and
$ds^2=-dT_1^2-dT_2^2+dX_1^2=-4dUdV/(1+UV)^2$.
For $U>0$, we also define $(T_1',T_2',X_1')$ priming all variables in Eq.(\ref{R21emb}).
In terms of $(T_1,T_2,X_1)$, the geodesic distance $d_\textrm{RM}$ between R and M  is simply written as
\begin{equation}
 \cosh d_\textrm{RM}=T_1(R)T_1(M)+T_2(R)T_2(M)-X_1(R)X_1(M)\ .
\end{equation}
From $(T_1, T_2, X_1)|_M=(V_M,1,V_M)$, we have
\begin{equation}
 \cosh d_\textrm{RM}=\frac{1-U_RV_R+2U_R V_M}{1+U_R V_R}=\frac{1}{az}+\left(\frac{1}{az}+1\right)U_R V_M\ .
\end{equation}
At the second equality, we eliminate $V_R$ using last expression in Eq.(\ref{txz_before}) and,
thus, $z$ appears in this expression. By the same way, the geodesic distance between L and M  is
\begin{equation}
 \cosh d_\textrm{LM}=\frac{1}{az}+\left(\frac{1}{az}+1\right)U'_L(V_M+\gamma)\ .
\end{equation}
Minimizing $d=d_\textrm{LM}+d_\textrm{RM}$ with respect to $V_M$, we obtain
\begin{equation}
 d=2\ln\left(\frac{2}{az}\right)+\ln\left[-\frac{(U'_L-U_R-\gamma U'_L U_R)^2}{4U'_L U_R}\right]+\mathcal{O}(z)\ .
\end{equation}
At AdS boundaries, $U_R$ and $U_L'$ relate to proper times $\tau_R$ and $\tau_L$ of
 quark and antiquark as
\begin{equation}
 U_R=-e^{-a\tau_R}\ ,\quad
 U_L'=e^{a\tau_L}\ .
\end{equation}
Therefore, we obtain
\begin{equation}
 d^\textrm{reg}=
  2\ln\left[\cosh \left\{\frac{a}{2}(\tau_L+\tau_R)\right\}+\frac{\gamma}{2}\exp\left\{\frac{a}{2}(\tau_L-\tau_R)\right\}\right]\ .
\label{dreg}
\end{equation}
where $d^\textrm{reg}\equiv [d-2\ln(2/az)]_{z=0}$.
The distance becomes longer as $\gamma$ increases.
Fig.\ref{geodesic_worldsheet} shows the geodesic for $\gamma=1,2,3$
in the target space coordinates $(t,x,z)$.
We can see how the geodesic distance becomes large as $\gamma$ increases from this figure.
We take the proper times as $\tau_R=\tau_L=0$.
Without the perturbation, origins of the proper times are chosen so that $t|_{\tau_R=0}=t|_{\tau_L=0}=0$.
However, because of the effect of the perturbation, 
the antiquark is slightly accelerated and has time shift.
Thus, we have $t|_{\tau_L=0}<0$ in the figure.
The geodesic is stretched between the quark and antiquark along the ``waist'' of the string worldsheet.
As $\gamma$ increases, the effect of the perturbation becomes significant and the waist becomes thicker.
In next section, we will find that this can be regarded as a visualization of the decay of the correlation
between the quark and antiquark.

\section{Fast scrambling}
\label{sec:FS}

\subsection{Correlation function}
\label{corr}

As the measure of the entanglement between quark and antiquark,
we consider the correlation function of forces acting on quark and antiquark.
The mutual information can be another measure,
which is computed from the entanglement entropy.
However, the holographic estimation of the entanglement entropy
is unclear for the dynamical probe string.
So, we only focus on the correlation function in this paper.
Since we gave a perturbation on the holographic EPR pair,
the background thermofield double state~(\ref{TFD}) would be changed as
$|\Psi\rangle \to W_L(\tau_0)|\Psi\rangle$
where $W_L(\tau_0)$ is the operation of the tiny change of the acceleration of the antiquark at the proper time $\tau=\tau_0$.
Therefore, the correlation function between quark and antiquark would be written as
\begin{equation}
 \langle F_L(0) F_R(0)\rangle_W \sim \langle \Psi|W_L^\dagger(\tau_0)F_L(0) F_R(0) W_L(\tau_0)|\Psi\rangle\ .
  \label{WFFW}
\end{equation}
Hereafter, we will sometimes omit arguments of operators for $\tau=0$, e.g. $F_L(0)=F_L$.

\begin{figure}
  \centering
  \subfigure[$\delta a>0$]
 {\includegraphics[scale=0.4]{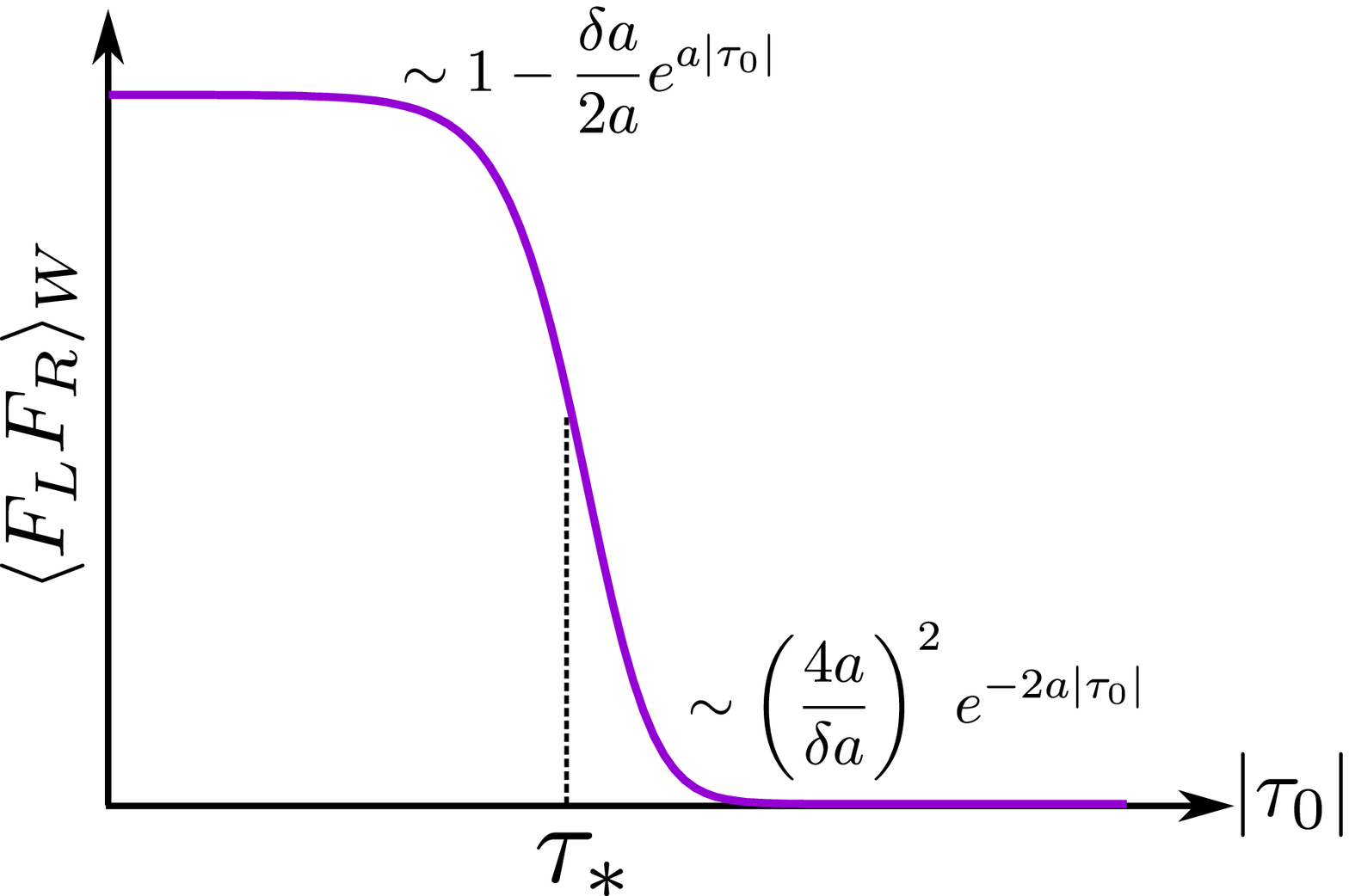}\label{o1}
  }
  \subfigure[$\delta a<0$]
 {\includegraphics[scale=0.4]{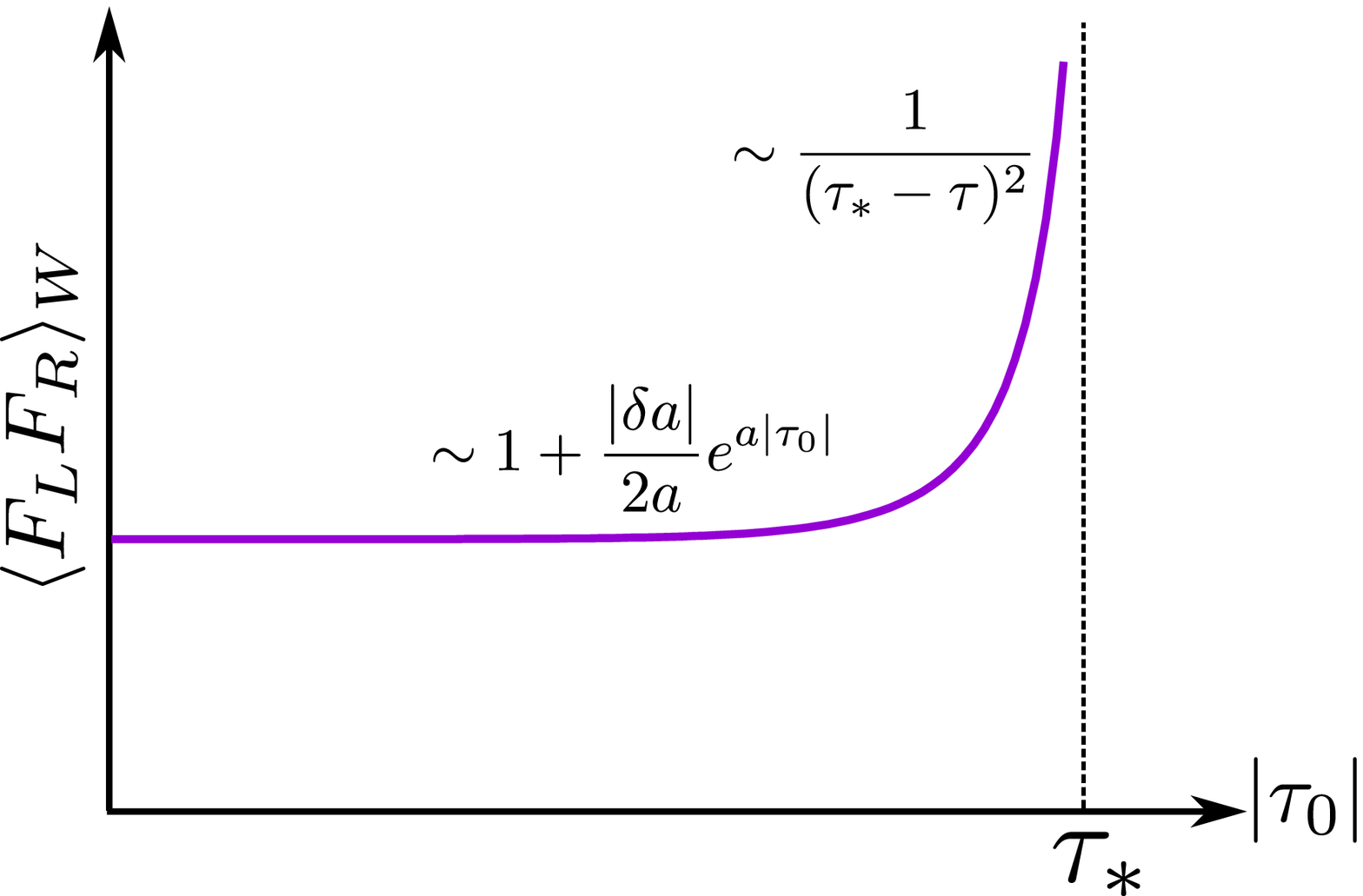}\label{o2}
 }
 \caption{
 Correlation function for $\delta a>0$ and $\delta a<0$.
 }
 \label{OTOCs}
\end{figure}

The conjugate bulk field for the force is $x(\tau,\sigma)$~\cite{Gubser:2006nz,Ishii:2015wua}.
This is a massless degree of freedom.
Differentiating the on-shell action of the string with respect to the boundary value,
we can compute the the two point functions.
Here, instead of following the traditional way, we use the geodesic
approximation to estimate the two point function:
\begin{equation}
 \langle F_L F_R\rangle_W \sim e^{-\Delta d}\ .
\label{geo_app}
\end{equation}
where $\Delta$ is the conformal weight and $d$ is the geodesic distance between AdS boundaries.
This approximation is effective for large $\Delta$.
The conformal weight for the string perturbation $x(\tau,\sigma)$ is $\Delta=1$.
We will use the geodesic approximation
just as a rough estimation of the correlator.\footnote{
For the pure AdS spacetime,
the geodesic approximations give exact results for any $\Delta$~\cite{Balasubramanian:1999zv}.
}
To obtain massive fields ($\Delta\gg 1$),
we can consider the D$p$-brane in AdS$_5\times S^5$ instead of the F-string.
Wrapping a subspace of $S^5$ in the $(p-1)$-dimensional part of the D-brane, we obtain a ``string'' in AdS$_5$.
We have a tower of massive fields on the D-brane as Kalza-Klein modes.\footnote{
We thank Tadashi Takayanagi for suggesting this idea.
}
Substituting Eq.(\ref{dreg}) into above expression, we obtain
\begin{equation}
 \langle F_L F_R\rangle_W \sim \left(1+\frac{\gamma}{2}\right)^{-2}=\left(1+\frac{\delta a}{4a}e^{a |\tau_0|}\right)^{-2}
  \ .
  \label{FF}
\end{equation}
Note that $\tau_0$ takes large negative value because of the limit~(\ref{SSlimit}).
We show the typical time dependence of the correlation for $\delta a>0$ in Fig.\ref{o1}.
(We will discuss the case of $\delta a<0$ in section.\ref{deca}.)
The effect of the perturbation becomes significant for $|\tau_0|> \tau_\ast$ where $\tau_\ast$ is the scrambling time: 
\begin{equation}
 \tau_\ast \sim \frac{1}{a}\ln\left(\frac{a}{\delta a}\right)\ .
\end{equation}
From Eq.(\ref{Unruh}) and the second equation of Eq.(\ref{ES}),
we have $a\propto E$. Thus, we obtain $\delta a/a=\delta E/E$.
From Eq.(\ref{ES}), we also obtain $E\sim S/\beta$. 
Therefore, the scrambling time is rewritten as
\begin{equation}
 \tau_\ast\sim \frac{\beta}{2\pi} \ln \left(\frac{S}{\beta \delta E}\right) \ .
\end{equation}
If we assume $\beta \delta E=\mathcal{O}(1)$, 
we obtain $\tau_\ast\sim \beta/(2\pi) \ln S$.
The holographic EPR pair is a fast scrambler.

As mentioned in Ref.\cite{Roberts:2014ifa}, 
the correlation function~(\ref{WFFW}) relates to 
the out-of-time-order correlator (OTOC) introduced
in Refs.\cite{Larkin,Maldacena:2015waa} as followings.
For an operator $\mathcal{O}$, we define the 
$\mathcal{O}_L=\mathcal{O}\otimes 1$ and $\mathcal{O}_R=1 \otimes \mathcal{O}^T$.
Substituting the explicit expression of the thermofield double state~(\ref{TFD}),
we rewrite the correlation function~(\ref{WFFW}) as 
\begin{equation}
\begin{split}
&\langle \Psi|W_L^\dagger(\tau_0)F_L F_R W_L(\tau_0)|\Psi\rangle \\
=
&Z^{-1}\sum_{n} 
 \langle n|e^{-\beta H/2}Fe^{-\beta H/2}
 W^\dagger(\tau_0)\,F\,W(\tau_0)|n\rangle \\
=&\langle W^\dagger(\tau_0)\,F(0)\,W(\tau_0)\,F(i\beta/2) \rangle_\beta\ .
\end{split}
\end{equation}
where 
$\langle \cdots \rangle_\beta \equiv Z^{-1}\textrm{tr}[e^{-\beta H}\cdots]$.
The correlation function relates to the OTOC.
From Eq.(\ref{FF}), the OTOC behaves at early time as $\sim 1-\delta a/(4 a)e^{a|\tau_0|}$
Therefore, the Lyapunov exponent for the accelerated quark is
\begin{equation}
 \lambda_L = a = \frac{2\pi}{\beta}\ .
\end{equation}
This saturates the bound in Ref.\cite{Maldacena:2015waa}.

\subsection{Correlation in the laboratory frame}

So far, we took the proper time of the quarks to express the correlation.
There is the other natural time coordinate:
the target space time coordinate $t$ which would be regarded as the laboratory frame.
In the limit of Eq.(\ref{SSlimit}), the relation between the proper time and target space time coordinate is given by
\begin{equation}
 t_L=-\frac{\gamma}{a}+\frac{1}{a}\sinh a \tau_L\ ,\quad
  t_R=\frac{1}{a}\sinh a \tau_R\ .
\end{equation}
Let us focus on the correlation of $t=0$ slice.
Then, we have $\tau_L=a^{-1}\sinh^{-1} \gamma$ and $\tau_R=0$ from above expressions.
Substituting them into Eq.(\ref{dreg}),
we can express the geodesic distance between two AdS boundaries.
The correlation function~(\ref{geo_app}) is given by
\begin{equation}
\langle F_L(t_L=0) F_R(t_R=0) \rangle_W=
\frac{4}{(1+\sqrt{1+\gamma^2})^2(\gamma+\sqrt{1+\gamma^2})}\ .
\label{FFtarget}
\end{equation}
The time for the shock injection is written as $t_0\equiv t_L|_{\tau_L=\tau_0}\simeq -\gamma/\delta a$.
Thus, we have $\gamma=\delta a |t_0|$.
We cannot find any exponential behaviour in Eq.(\ref{FFtarget}) in terms of $t_0$.
The scrambling time can be estimated as $\gamma \sim 1$, i.e. 
$|t_0|\sim 1/\delta a \sim \beta (E/\delta E) \sim S/\delta E$. We do not find
the fast scrambling in the the laboratory frame.

\subsection{Decreasing the acceleration}
\label{deca}

We have considered the case of $\delta a>0$.
Here, we focus on the decreasing of the acceleration $\delta a<0$.
For the negative $\delta a$, it is remarkable that 
the correlation~(\ref{FF}) blows up at $|\tau_0|=a^{-1}\ln(2a/|\delta a|)\equiv \tau_\ast$.
Fig.\ref{o2} shows the time dependence of the correlation for $\delta a<0$.

\begin{figure}
\begin{center}
\includegraphics[scale=0.6]{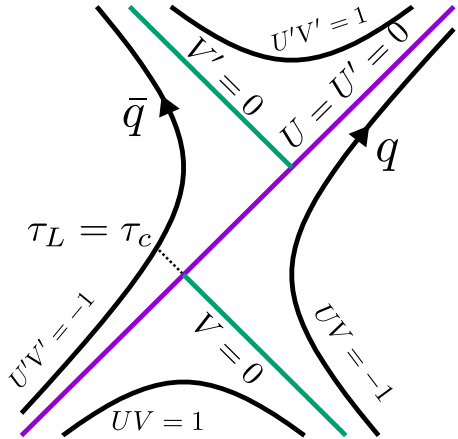}
\end{center}
\caption{
 Spacetime structure of the perturbed string worldsheet for the decreased acceleration.
 An ``one-way traversable wormhole'' is created after $\tau_L=\tau_c$.
}
\label{shock_negative}
\end{figure}

To see the origin of the divergence,
we consider the spacetime structure of the string worldsheet.
Since the shift of the $V$-coordinates at the shock surface~(\ref{Vshift}) becomes opposite, the spacetime structure of the worldsheet becomes like as in Fig.\ref{shock_negative}.
For $\tau_L>\tau_c$, 
the quark and antiquark are causally connected, where 
$\tau_c$ is shown in the figure. (
Explicitly, we can write $\tau_c=-|\tau_0|+a^{-1}\ln(2a/|\delta a|)$.)
The causal connection is one-way: one can send a signal from the quark to antiquark,
but the opposite is causally forbidden.
In that sense, the initial non-traversable wormhole becomes the one-way traversable wormhole.
Points on left and right boundaries can be connected by a light-like geodesic.
Then, the geodesic distance becomes zero and this is the origin of the divergence of the correlation function.

There is the relation between the correlation function and OTOC only for $\tau_L<\tau_c$.
For $\tau_L>\tau_c$, the left CFT is disturbed by the right CFT. Hence, the left Hamiltonian would be written as $H_L+\theta(\tau_L-\tau_c)H_{LR}(\tau_L)$ where $H_{LR}$ is the interaction term between left and right CFTs while the right Hamiltonian is unchanged.
Thus, $|n\rangle_L$ cannot be regarded as an eigenstate for $\tau_L>\tau_c$.
In other words, Eq.(\ref{FF}) is regarded as the OTOC only before the divergence.
This suggests that, depending on inserted operators, the OTOC can diverge within finite time.
In Ref.\cite{Gao:2016bin},
it has been proposed that a traversable wormhole can be created
by the double trace deformation.
Our results would suggest another way to
create the traversable wormhole using the EPR pair.

 \section{Two shocks}
 \label{sec:two}

We have changed the acceleration of the antiquark just once:
For $\tau>\tau_0$, the acceleration is eternally deviated.
This may seem unnatural operation. 
In this section, we will consider the ``undo'' of the operation: $a\to a' \to a$.
Explicitly, we take the following forms of $f_t$ and $f_x$:
\begin{equation}
\begin{split}
&f_t(\tau)=
\begin{cases}
\frac{1}{a}\sinh a \tau & (\tau\leq \tau_0)\\
 \frac{1}{a'}\sinh[a'\tau+c_1]+c_2\ & (\tau_0<\tau\leq \tau_1)\\
 \frac{1}{a}\sinh[a\tau+c_3]+c_4\ & (\tau>\tau_1)
 \end{cases}\ ,\\
 -&f_x(\tau)=
\begin{cases}
 \frac{1}{a}\cosh a \tau & (\tau\leq \tau_0)\\
 \frac{1}{a'}\cosh[a'\tau+c_1]+c_2'\ & (\tau_0<\tau\leq \tau_1)\\
 \frac{1}{a}\cosh[a\tau+c_3]+c_4'\ & (\tau>\tau_1)
 \end{cases}
\ .
\end{split}
\label{fspert}
\end{equation}
where
\begin{equation}
\begin{split}
&c_3=(a'-a)(\tau_1-\tau_0)\ ,\quad
 c_4=\left(\frac{1}{a}-\frac{1}{a'}\right)
 \left\{\sinh a\tau_0-\sinh[a'\tau_1-(a'-a) \tau_0]\right\}\ ,\\
&c_4'=\left(\frac{1}{a}-\frac{1}{a'}\right)
 \left\{\cosh a\tau_0-\cosh[a'\tau_1-(a'-a) \tau_0]\right\}\ ,
\end{split}
\end{equation}
and $c_1$, $c_2$ and $c_2'$ are defined in Eq.(\ref{cc}).
These constants are determined from the continuity of above functions and their derivatives.
Substituting above expressions into Eq.(\ref{Mdef}), we have a rectangular function:
\begin{equation}
 M(\tau)=a^2+(a'{}^2-a^2)\theta(\tau-\tau_0)\theta(\tau_1-\tau)\ .
\end{equation}
Note that, in case of the Vaidya spacetime, we cannot ``undo'' the mass change 
if we impose the null energy condition for the infalling matter.

For $\tau<\tau_0$ and $\tau_0<\tau<\tau_1$, we use $(U,V)$- and $(U',V')$-coordinates
defined in Eqs.(\ref{UV_usigma}) and (\ref{UpVp_usigma}), respectively.
For $\tau>\tau_1$, we also introduce $(U'',V'')$-coordinates as
\begin{equation}
 U''=e^{a\tau} \ , \quad V''=-\frac{\sigma-a}{\sigma+a}e^{-a\tau} \quad \Longleftrightarrow\quad 
  \tau=\frac{1}{a}\ln U''\ ,\quad \sigma=a\, \frac{1-U''V''}{1+U''V''}\ .
\label{UppVpp_usigma}
\end{equation}
By the similar way as the derivation of Eq.(\ref{Vshift}),
the matching condition at the second shock surface $\tau=\tau_1$ is given by
\begin{equation}
 V''=V'-\gamma e^{-a(\tau_1-\tau_0)}\ ,
 \label{Vshift2}
\end{equation}
where we took the double scaling limit~(\ref{SSlimit}) keeping $\tau_1-\tau_0$ fixed.
In this limit, the first and second shock surfaces degenerate: the first shock is at $U=U'=0$ and the second shock is at $U'=U''=0$. The matching condition for $V$ and $V''$-coordinates as
\begin{equation}
 V''=V+(1-e^{-a(\tau_1-\tau_0)})\gamma\ .
\end{equation}
We can obtain the correlation function for holographic EPR pair with two shock just by replacing $\gamma\to (1-e^{-a(\tau_1-\tau_0)})\gamma$ in the single shock case. Since we gave two shocks at $\tau=\tau_0, \tau_1$, the thermofield double state~(\ref{TFD}) would be perturbed as $|\Psi\rangle \to W_L'(\tau_1)W_L(\tau_0)|\Psi\rangle$.
The correlation function is given by
\begin{equation}
 \langle F_L F_R\rangle_{WW'} \sim \left(
1+\frac{\delta a}{4a}e^{a |\tau_0|} (1-e^{-a(\tau_1-\tau_0)})\right)^{-2}
  \ ,
  \label{FF2}
\end{equation} 
where $\langle\cdots \rangle_{WW'}$ represents the expectation value of the $W_L'(\tau_1)W_L(\tau_0)|\Psi\rangle$.
We again obtain $\lambda_L=a=2\pi/\beta$.
Assuming that the time scale of the change of the acceleration as
$\tau_1-\tau_0\sim T \sim a$, we have $1-e^{-a(\tau_1-\tau_0)}=\mathcal{O}(1)$.
Therefore, we also have the fast scrambling result even for the two shock case.

\section{Conclusion}

We have studied scrambling behaviour of the holographic EPR pair.
The gravity picture of the holographic EPR pair
is the fundamental string whose endpoints 
are accelerated in opposite direction.
The worldsheet metric
is given by an eternal AdS$_2$ black hole geometry.
It follows that quarks are in thermofield double state.
As a perturbation,
we slightly increased the acceleration of the antiquark.
If we give the perturbation at a sufficiently early time,
its effect becomes significant even for a tiny change of the acceleration.
It quickly destroyed the correlation between the quark and antiquark.
Its time scale is given by $\tau_\ast\sim \beta\ln S$. 
This indicates that the holographic EPR pair is a fast scrambler.
From the early time behaviour of the correlation function,
we also estimated the Lyapunov exponent, $\lambda_L=2\pi/\beta$.
This saturates the bound proposed in Ref.\cite{Maldacena:2015waa}.
The gravity dual of the EPR pair is not Einstein gravity but the probe string.
Our results suggest that the fast scrambling behaviour or
the saturation of the Lyapunov bound 
do not directly imply the existence of a dual Einstein gravity.
We also slightly decreased the acceleration of the antiquark.
Then, the quark and antiquark are causally connected.
In the worldsheet point of view, 
the one-way traversable wormhole is created.
Two points in boundaries can be lightlike separated and this cause the 
the divergence of the correlation function.
We also studied the two shock case: The acceleration was changed as $a\to a'\to a$.
We again found the fast scrambling and the saturation of the Lyapunov bound.

We still have some issues and future works to be addressed.
For the estimation of the correlation between the quark and antiquark,
we used the geodesic approximation~(\ref{geo_app}).
For more precise estimation,
we need to compute it from the onshell action.
It would be nice if we can do that following the method in Ref.~\cite{Gubser:2006nz,Caceres:2010rm}.

In the holographic EPR pair, the quark and antiquark are causally disconnected from the beginning.
One can also consider more realistic setup: Dynamical creation of the event horizon on the worldsheet.
It 
has been demonstrated in Refs.\cite{Chernicoff:2013iga,Ishii:2015qmj}.
It is interesting to study the scrambling of quarks for the dynamical horizons.

For the computation of the correlation in this work and also in Ref.\cite{Shenker:2013pqa},
the action of the system did not play important role.
It is just given by the geometrical information, i.e, geodesics.
In recent years, the quantum complexity is regarded as an important quantity
to explore the property of the black hole interior~\cite{Susskind:2014rva,Stanford:2014jda}.
It has been conjectured
that the complexity is dual to the action of the Wheeler-DeWitt patch~\cite{Brown:2015bva,Brown:2015lvg}.
The holographic EPR pair would give a simple model to study the complexity.
For the evaluation of the quantum complexity,
not only the geometry but also the action is important.
The Einstein gravity and Nambu-Goto action would
give a qualitative difference in the complexity.

\ \\
\noindent
\textbf{Note Added:}\\
When this manuscript was prepared for submission,
we noticed a related work by Shinji Hirano and Nilanjan Sircar~\cite{Hirano}.
The independent work~\cite{deBoer:2017xdk} by J.~de Boer, E.~Llabr\'{e}s, J.~F.~Pedraza and D.~Vegh appeared almost simultaneously,
which studies the related subject to ours.

\acknowledgments

The author thanks Pawel Caputa, Shinji Hirano, So Matsuura and Tadashi Takayanagi for valuable discussions and comments.
The author got the idea for this work while at ``Workshop on OTO correlators'' held in Osaka University.
The work was supported by JSPS KAKENHI Grant Number 15K17658.

\end{document}